\documentstyle[pre,aps,psfig]{revtex}

\begin{document}

\title{Cell theory of the electroosmosis in the concentrated diaphragms
consisting of spherical particles of colloid dispersity with the account
of the insoluble boundary layer influence}
\author{
V. N. Shilov \footnote{e-mail: shilov@bioco.kiev.ua}
and  N. I. Zharkikh
}
\address{ F. D. Ovcharenko Institute of Biocolloidal Chemistry,
National Academy of Sciences of Ukraine,\\ 42 Vernadsky blvd.,
252142 Kyiv, Ukraine}
\draft
\date{\today}
\maketitle
\begin{abstract}

The electroosmotic transfer (ratio of velocity of liquid to electric
current density) and conductivity of disperse system were calculated
as functions of volume fraction of disperse particles. The considered
model of electric double layer (EDL) was generalized by taking into
account the reducing dissolving ability of liquid near the surface
(non-dissolving boundary layer). The problem was solved for limiting
case of large degree of EDL overlapping in interparticle space. The
obtained results explain the main features of experimental data.

\end{abstract}


\section{Premises for the problem solution and the model characteristics}

The achievements in the quantitative theory of the suspension kinetic
properties concern almost exclusively diluted systems when the consideration
of a single disperse particle in a infinite disperse medium volume can be in
the theory base. In the concentrated systems (pastes, diaphragms being from
colloid particles) the picture of the phenomena is becoming complicated
substantially under the influence of the interaction between equilibrium and
polarization fields that strongly impedes the theory construction. However,
just in the concentrated systems with the particles of the colloid
dispersity the quite interesting influence on the kinetic properties in
anomalous fluid layers near the surface, inasmuch as in the concentrated
systems these layers occupy substantial volume part of the threshold space.

Just in such systems the authors [1-3] systematically were investigating the
electroosmosis as a function of the disperse phase concentration and had
found simple but enough general regularities: such as the linear dependence
of the electroosmotic transfer $P_e$ a reverse value of the particle
concentration, the part of the segment being cut off by the straight line
$P_e(C^{-1})$ on the abscissa with the fluid boundary layer thickness, other
peculiarities. For the explanation of these regularities the model of the
near wall layer of the fluid in which it combines the non-dissolving volume
properties with the hydrodynamic mobility was proposed in the work [4]. The
calculations in the work [4] were based on the simplest geometric model
representing the porous space of the suspension by plane capillaries. The
model of the plane capillary, however, for suspension porous space is too
rough, all the more, the question is the effects substantially depending on
the geometry interface (in particular the dependence of $P_e$ on the volume
particle fraction). Therefore, the question arised is it possible to explain
the experiment in the model more realistically than plane capillaries taking
into consideration the particle surface curvature by the model of the
hydrodynamically mobile non-dissolving near wall layer.

The possibility to receive the answer to this question has appeared in
connection with the development [5,6] of the procedure for the use of the
cell model widely applied while solving purely hydrodynamic tasks to the
electrokinetic problems.

\section{Statement of a task}

In the cell model the consideration of a suspension is the consideration of
one of its typical structural unit - $a$ radius of a particle ``belonging'' to
it in the fluid volume. And the main simplification consists in that the
boundary of this volume is considered to be a sphere (instead of
corresponding polyhedron). The fraction of the volume being occupied by a
particle in the cell corresponds to the volume particle fraction in the
suspension. Having designated the cell radius by $c$ we obtain:
\begin{equation}
\alpha =a^{3}/c^{3}.  \eqnum{1}
\end{equation}
Introducing the idea about non-dissolving boundary layer we shall designate
the radius of the dissolving and non-dissolving layer in the cell by $b$ .
Then the thickness $h$ of the non-dissolving boundary layer will be equal:
\begin{equation}
h=b-a,  \eqnum{2}
\end{equation}

and the volume fraction of the non-dissolving water (see Fig. \ref{fig1})
\begin{equation}
\alpha _{1}=(b^{3}-a^{3})/c^{3}.  \eqnum{3}
\end{equation}

The presence of the non-dissolving layer means that the diffuse shell of the
double electric layer is completely in the dissolving water. Namely, the
volume force is applied to it from the electric field side. We'll restrict
our consideration about high degree of overlapping of diffuse double layer of
neighbor particles, when Debye length $R_D$ is much larger that thickness of
dissolving layer: $R_D/(c \times b-h)<<1$. It corresponds to the
supposition that the volume charge density and also the volume force are
constant in all points of the dissolving layer (homogeneous approximation).
Then for the calculation of the electroosmosis rate of the 1-1 valent binary
electrolyte the following equation system must be solved:

\begin{equation}
-\eta \;rot\,\/rot\;\vec{\nu}-grad\,(p+RT\gamma ^{+}\mu ^{+}+RT\gamma
^{-}\mu ^{-})=0,  \eqnum{4}
\end{equation}
\begin{equation}
div\vec{\nu}=0,  \eqnum{5}
\end{equation}
\begin{equation}
div\,grad\,\mu ^{\pm }=0,  \eqnum{6}
\end{equation}
where $\eta$ is the fluid viscosity, $p$ is the pressure, $\nu$ is the fluid rate, is
dimensionless (in the $RT$ units) electrochemical ion potentials, $C$ are their
concentrations. In this case $\gamma{^\pm}= C - C_o$, $C_o$ is of the concentration of the
solution being equilibrium with the suspension:
\begin{equation}
c^{\pm } =\mp \frac{G}{2} +\sqrt{\/\frac{G^{2} }{4} +C_{0}^{2} },
\eqnum{7}
\end{equation}
\begin{equation}
\rho =F\,(\,C^{+} -C^{-} )\equiv -\,FG,
\eqnum{8}
\end{equation}
where $F$ is the Faraday constant.

The particle charge $Q$ is expressed through $G$ :
\begin{equation}
Q=-\frac{4\pi }{3} (\,C^{3} -b^{3} )\/\/FG.
\eqnum{9}
\end{equation}

In the non-dissolving layer the volume force does not act on the fluid
because on the definition the ions do not penetrate there. For this region
of equation system (4) - (6) is simplified at the expense of all members
containing .

(4) - (6) system must be solved jointly with the edge conditions:
\begin{equation}
\left. \nu _{r} \right| _{\,r=a} =0,
\eqnum{10}
\end{equation}
\begin{equation}
\left. \nu _{\theta } \right| _{\,r=a} =0,
\eqnum{11}
\end{equation}
\begin{equation}
\left. \nu _{r} \right| _{\,r=b-0} =\left. \nu _{r} \right|_{\,r=b+0},
\eqnum{12}
\end{equation}
\begin{equation}
\left. \nu _{\theta } \right| _{\,r=b-0} =\left. \nu _{\theta } \right|
_{\,r=b+0},
\eqnum{13}
\end{equation}
\begin{equation}
\left. P\,\right| _{\,r=b-0} =\left. P\,\right| _{\,r=b+0},
\eqnum{14}
\end{equation}
\begin{equation}
\left. \Pi _{r\/\theta } \right| _{\,r=b-0} =\left. \Pi _{r\/\theta }
\right| _{\,r=b+0},
\eqnum{15}
\end{equation}
\begin{equation}
\left. rot\,\nu \/\right| _{\,r=c} =0,
\eqnum{16}
\end{equation}
\begin{equation}
\left. (-D^{\pm } \frac{\partial \/\mu }{\partial \/r} +\nu _{r} )\right|
_{\,r=b+0} =0.
\eqnum{17}
\end{equation}

Here $\Pi$ is the tangent component of the tensor of the viscous stresses
(during its calculation shall take the same viscosity for both the
dissolving and non-dissolving layers) the symbols $b\pm o$ designate the limits
of the corresponding values while the radius tends to $b$ both from the right
and from the left respectively. $D^\pm$ is the ion diffusion coefficients.
Condition (16) is the cell edge condition for the Cuvabara model [7].

Besides, the conditions characterizing the electroosmosis process must be
given:

\begin{equation}
\left. \frac{F}{\cos \theta }(-D^{+}C^{+}\frac{\partial \mu ^{+}}{\partial r}%
+D^{-}C^{-}\frac{\partial \mu ^{-}}{\partial r}+\nu
_{r}(\,C^{+}-C^{-}))\/\right| _{\,r=c}=I_{E},
\eqnum{18}
\end{equation}

\begin{equation}
\left. \mu ^{-}+\mu ^{+}\right| _{\,r=c}=0,
\eqnum{19}
\end{equation}

\begin{equation}
\left. \int_{r=a}\Pi
ds+RT\int_{b<r<c}(C^{+}-C^{-})grad(\mu^{+}-\mu^{-})d\nu=0, \right.
\eqnum{20}
\end{equation}
where $I_E$ is electric current, $\Pi$ is the tensor of the viscous
stresses. Condition (19) means the absence of the concentration
difference in the suspension. Condition (20) means the compensation of
the viscous force acting a particle by sum of the volume electrical
forces acting the charged fluid. The absence of the difference in
pressure on the diaphragm made from particles.

The volume rate $I_v$ of the fluid in the cell model is expressed by the formula
\begin{equation}
I_{V} =\left. \frac{\nu _{r} }{\cos \theta } \right| _{\,r=c},
\eqnum{21}
\end{equation}
and the average electric field E is equal
\begin{equation}
E=\left. \frac{RT}{Fc\,\cos \theta } (\mu ^{+} -\mu ^{-} )\,\right|
_{\,r=c}.
\eqnum{22}
\end{equation}

Having determined these values we have found the electroosmotic
transfer $P_e$ and electroconductivity $K$ :
\begin{equation}
P_{e} =I_{V} /I_{E},\eqnum{23}
\end{equation}
\begin{equation}
K_{\Sigma } =I_{E} /E.\eqnum{24}
\end{equation}

\section{Problem solution}

The edge problem (4) - (20) is reduced to 12 linear algebraic equations of
the following by substitution of the general solutions:
\begin{equation}
\mu ^{\pm } =\cos \theta \,(A_{1}\pm r+A_{2}^{\pm } r^{-2} ),\eqnum{25}
\end{equation}
\begin{equation}
\nu _{r} =\cos \theta \,(B_{1} -\frac{2B_{2} }{r^{3} } +\frac{B_{3} r^{2} }
{10} +\frac{B_{4} }{r} ),\eqnum{26}
\end{equation}
\begin{equation}
\nu _{\theta } =-\sin \theta \,(B_{1} +\frac{B_{2} }{r^{3} } +\frac{B_{3}
r^{2} }{5} +\frac{B_{4} }{2r} ),\eqnum{27}
\end{equation}
\begin{equation}
p=\frac{\eta \,\cos \theta }{a} (B_{3} r+\frac{B_{4} }{r^{2} } ).\eqnum{28}
\end{equation}

The solution for the rate at $b \le r \le c$ formally coincides with (27) if one
considers the instead the pressure.
\begin{equation}
S=p+RT\gamma ^{+} \mu ^{+} +RT\gamma ^{-} \mu ^{-}.\eqnum{29}
\end{equation}

The calculation of the integrals included in condition (20) using (25) -
(28) shows that (20) is equal to the following
\begin{equation}
B_{4} -\frac{a^{2} \,RT}{3\eta } (\gamma ^{+} A_{2}^{+} +\gamma ^{-}
A_{2}^{-} )(\,\frac{1}{a} -\frac{b^{3} }{a^{3} } )=0.
\eqnum{30}
\end{equation}

Rest edge conditions are transformed into the linear equations in trivial
way.

The analytical expressions obtained for $P_e$ and $K_\Sigma$ from the solution of these
equations are quite awkward and they are inconvenient. Therefore, the
solution analysis will be based on the numerical investigation of the most
typical situations.

While concretizing the parameters of equations (4) - (20) we gave their
values characteristic for the systems being studied in the experiments on
the electroosmosis on the diaphragms based on the synthetic diamond with the
particles of 150 nm size in the HCl solutions with the 0.01 - 1.0 mmole/l
concentration. The viscosity was 1.0 mPa.c. The diffusion coefficients were
$D^+ = 8.86\cdot 10^{-5}$ cm${^2}$/c, $D^-=1.94\cdot 10^{-5}$ cm${^2}$/c. The dependencies of the
electroosmositic transfer $P_e$ on the particle charge, on the volume fraction
and the non-dissolving boundary layer thickness calculated using the
mentioned values of the system characteristics are shown in Figs.
\ref{fig5}-\ref{fig7}.

\section{The calculation result discussion}

The main circumstance while interpreting the dependencies in Figs.
\ref{fig5}-\ref{fig7} is
that the local conductivity $K$ and the charge density $\rho$ in the dissolving fluid
of the threshold space in the homogeneous approximation has the dependence
character depicted in Figures \ref{fig2} and \ref{fig3}.

If the average field strength applied
to the suspension is not changed the electroosmosis rate is proportional to
the particle charge and the volume charge density ($I_v Q \rho$ ). The fluid in the
porous space is charged, therefore, the flow gives birth to the convection
component of the electric current $I_v(Q) = I_v(Q)/I_E/$. On the other hand, the
migration component of the ekectroconduction associated with the ion
movement through the fluid is proportional to the local conductivity. Thus,
at small charges the sum current will be determined mainly by the
electromigration component and must slowly grow with the charge increase
(Fig. \ref{fig4}, region 1). At great particle charges the sum is determined by the
convection component and is proportional to the charge square (Fig. \ref{fig4},
region 2).

With the account of the mentioned statement it is not difficult to explain
the maxima on the dependencies $P_e(Q) = I_v(Q)/I_E(Q)$ in Fig. 5: region 1
corresponds to the ascending branch, region 2 corresponds to the descending
branch in Fig. \ref{fig4}. On appearing of the non-dissolving layer the volume charge
moves away from the surface. In this case, inasmuch as the particle charge
does not change, the sum volume force remained as previous one, only the
region of its application also moved away from the surface, i.e., to there
where the flow is developing easily. Therefore, also the electroosmosis rate
and the convection current grows and the latter increases more
substantially, inasmuch as it is determined by the movement only in the
region of the dissolving volume where the local rate is high, then as the
flows in the non-dissolving volume being retarded by the surface make the
contributions in the integral electroosmosis rate. Therefore, at small
charges (region 1 in Fig. \ref{fig4}) when the current is determined by the
electromigration component and does not increase with the growth in the
non-dissolving volume; the electroosmotic transfer is growing with the
average electroosmosis rate. At great charges (region 2, Fig. \ref{fig4}) when the
current is determined by the convection component and is growing with the
non-dissolving volume more rapidly than the electroosmosis rate does the
electroosmotic transfer drops (Fig. \ref{fig6}). The increase in the electroosmotic
transfer at simultaneous sharp increase of the electroconductivity observed
by the authors [8] in the diamond diaphragms during the destruction of the
non-dissolving boundary layer under the action of the saccharose solution
being introduced in the diaphragm corresponds to this. As to the dependence
of the electroosmotic transfer on the reverse value of the volume fraction
as it is seen from Fig. \ref{fig7} the linear dependencies $P_e(\alpha{^-1})$ are
obtained in wide interval of the volume fraction values including even
at the relatively its small values when the porous space cannot be
considered completely in the homogeneous approximation.

Thus, the calculations of the electroosmotic transfer by the cell model
based on the ideas about the near wall layer as about the non-dissolving and
hydrodynamically mobile layer show the complete correspondence of the
qualitative regularities obtained by the authors. This allows to hope for
the receiving a quantitative information about the near wall layer while
comparing in details the calculations obtained with the experimental data.

\section*{Acknowledgments}

The authors are grateful to Alexejev O. and Bojko Y. for useful
conversations.

\newpage
\begin{figure}[tbp]
\caption{
Structural elements in the cell model of a disperse system.
$a$-radius of a particle; h-thickness of the non-dissolving layer;
$b$-radius of the non-dissolving layer in the cell; $c$-cell radius;
1-particle surface charge; 2-external boundary of the non-dissolving
layer; 3-external boundary of a diffuse layer.
}
\label{fig1}
\end{figure}

\newpage
\begin{figure}[tbp]
\caption{
Dependencies of the electroosmotic transfer,$P¥ÿ(cm^3/K)$ and diaphragm
electroconductivity $K,(1/ohm cm)$ on particles charge $Q$.
}
\label{fig5}
\end{figure}

\newpage
\begin{figure}[tbp]
\caption{
Dependencies of the electroosmotic transfer $Pe$ and diaphragm
electroconductivity $K$  on the thickness of the non-dissoling
boundary layer $h ({m^{-10}})$.
}
\label{fig6}
\end{figure}

\newpage
\begin{figure}[tbp]
\caption{
Dependencies of the electroosmotic transfer $P_e$ and diaphragm
electroconductivity $K$  on the reciprocal volume particle fraction
( different values of the non-dissolving boundary layer thickness $h,m^{-10}$).
}
\label{fig7}
\end{figure}

\newpage
\begin{figure}[tbp]
\caption{Dependence of local conductivity $K$  in the dissolving fluid on
particles charge $Q$ (homogeneous approximation).}
\label{fig2}
\end{figure}

\newpage
\begin{figure}[tbp]
\caption{
Dependence of charge density $p$ in the dissolving fluid on particles
charge $Q$ (homogeneous approximation).
}
\label{fig3}
\end{figure}

\newpage
\begin{figure}[tbp]
\caption{
Dependencies of convection $I_{ec}$ and migration $I_{em}$ components of the
electric current and the sum current $I_e$ on particles charge $Q$.
}
\label{fig4}
\end{figure}

\end{document}